\documentclass[epj]{svjour} 
\usepackage{latexsym}
\usepackage{graphics}
\usepackage{subfig}
\usepackage{amsmath,amssymb}

\begin{document}

\title{Prisoner's Dilemma: non-trivial results for the lowest temptation level in the Darwinian and Pavlovian evolutionary strategies}
\author{Marcelo Alves Pereira\inst{1,*} \and Alexandre Souto Martinez\inst{1,2}
}


\institute{Universidade de S\~{a}o Paulo - Faculdade de Filosofia, Ci\^{e}ncias e Letras de Ribeir\~{a}o Preto \\
Av. Bandeirantes, 3900, 14040-901 \\
Ribeir\~{a}o Preto, S\~{a}o Paulo, Brazil \\
$^*$marceloapereira@usp.br
\and 
National Institute of Science and Technology in Complex Systems \\
asmartinez@usp.br}

\date{Received: date / Revised version: date}

\abstract{
The lowest temptation level ($T = 1$) is considered a trivial case for the Prisoner's Dilemma.
Here, we show that this statement is true only for a very particular case, where the players interact with only one player.
Otherwise, if the players interact with more than one player, the system presents the same possible behaviors observed for $T > 1$.
In  the steady state, the system can reach the cooperative, chaotic or defective phases, when adopting the Darwinian Evolutionary Strategy and the cooperative or {\it quasi}-regular phases, adopting the Pavlovian Evolutionary Strategy.
} 

\maketitle

In the Prisoner's Dilemma (PD), two players confront themselves and each one can either cooperate ($C$) or defect ($D$).
Under mutual cooperation, both players receive a payoff $R$ (\emph{reward}).
If both are defectors, their payoff is $P$ (\emph{punishment}).
When a player cooperates and the other defects, they receive $S$ (sucker) and $T$ (\emph{temptation}), respectively~\cite{axelrod_1984}.
The dilemma appears under the conditions $T>R>P>S$ and $R>(S+T)/2$~\cite{axelrod_1984}.
In these circumstances, the best individual strategy is to defect, because it assures a higher payoff than cooperation, independently of the opponent decision (Nash equilibrium).
However, the best global strategy is cooperation.

When PD is played repeatedly, it is called Iterated Prisoner Dilemma (IPD).
A computer tournament was proposed by Axelrod in the 80's~\cite{axelrod_1980,axelrod_1980b} to compare different strategies playing the IPD.
The winner strategy was {\it tit-for-tat} (TFT), where a player cooperates in the first round and subsequently, copies the opponent last round action.
It has only one time step memory.
Cooperation emergence as a stable strategy made the PD popular~\cite{axelrod_1981}.

In a spatial PD game, all players play against their neighbors, they sum the payoffs and update their states.
The rules to update the states depend on the adopted strategy.
Here, we consider two deterministic strategies based on the total payoff: \emph{Darwinian Evolutionary Strategy (DES)}~\cite{nowak_1992,soares_2006} and \emph{Pavlovian Evolutionary Strategy (PES)}~\cite{fort_2005,kraines_1989,kraines_1993,kraines_1995,posch_1999,lorberbaum_2002,nowak_1993b}.
A player adopting the DES uses the strategy of copying the best adapted player state (fittest player) among the interacting players.
This behavior can be compared to the Darwin's natural selection principle, the ``survival of the fittest''~\cite{beyer_2002}.
The PES is a win-stay, lose-shift strategy~\cite{pereira_pavlov}.
Defined an aspiration level (AL)~\cite{posch_1999}, a player compares his/her total payoff to it.
If the total payoff is greater than AL, the player keeps his/her current state, otherwise switches it.
This behavior can be thought as: \textit{``never change a winning team''}.

The main variable in the PD problem is the \emph{temptation} payoff.
Same temptation values yield a total payoff to the players that force they to switch their states, which cause a phase transition.
These values of temptation are the critical temptation values and they depend on the adopted strategy, on the system connectivity and the neighborhood configuration~\cite{pereira_pavlov}.
Here, we only consider the particular case $T = 1$, the lowest temptation level to defect.
We choose the following payoffs: for DES: $R = 1$ and $P = S = 0$; and for PES: $R = 1$, $P = -R = -1$ and $S = -T = -1$.
The AL for the PES is to receive a positive payoff.
When a cooperator plays against a defector, if $T = 1$ and $R = T$, both have the same payoff.
Due to this result, a previous study~\cite{duran_2005} has considered, explicitly, $T = 1$ as a trivial case (other authors do not even mention this case), since players do not switch their states during time evolution.
Meanwhile, this statement is true only for players using the DES in the case of each player interacting with only one neighbor.
If players play with more than one neighbor, they can switch their states and in fact they do.
In the following we show that $T = 1$ is indeed a non-trivial case.

Now, consider players, located in the sites of a one-dimensional lattice (with periodic boundary conditions). Each player can cooperate or defect, and play the IPD with $z$ neighbors (coordination number).
If $z$ is even, the player interacts with $z/2$ players to the right-hand side and $z/2$ to the left one.
If $z$ is odd, player \emph{self-interacts} (plays against his/her own state) and with $(z-1)/2$ players on each side.
The system order parameter is the asymptotic proportion of cooperators, $\rho_\infty$.
This geometry has been studied in Ref.~\cite{pereira_pavlov,pereira_2008_IJMPC,pereira_2008_BJP} and gives the same results for critical temptation values as for regular lattices, once it only depends on the connectivity of the system.
It also allows to numerically explore the whole parameter space \cite{pereira_2008_BJP} which permitted us to detect a new phase ({\it quasi}-regular) for the Pavlov strategy~\cite{pereira_pavlov}.

The variable $\theta_i$ represents the $i$-th player state, if $\theta_i = 0$, the player defects and if $\theta_i = 1$, the player cooperates.
If a player interacts with more than one player, the total payoff (sum of all his/her payoffs per interaction) is given by~\cite{pereira_pavlov}:
$G_{i}^{(z;~c_i)}(\theta_i) = Tc_i + P(z-c_i)$, for $\theta_i = 1$, and 
$G_{i}^{(z;~c_i)}(\theta_i) = Rc_i + S(z-c_i)$, for $\theta_i = 0$.
The quantities $T$, $R$, $P$ and $S$ are the PD payoff values; $z$ is the system connectivity; and $c_i$ is the number of cooperators in the $i$-th player neighborhood.
During time evolution, players can organize themselves in well defined cooperative or defective clusters, which define the borders between them.
Clusters interact among themselves, establishing the invasion processes.
In a cooperative cluster, inner members have a higher payoff than in the defective ones, and the payoff differences along borders are more remarkable.
On one hand, for players adopting the DES, these payoff differences along the border region are essential to determine these player states switching due to the payoffs comparison.
Whereas, in large clusters, inner players do not switch their states, since every player has the same state and payoff. 
Cooperators/defectors invade defective/cooperative clusters from the cluster border players~\cite{pereira_2008_IJMPC}.
On the other hand, if they are adopting PES, the switching process can take place in any location of the cluster.
Inner players can switch their states, once they do not compare their payoffs with the neighboring ones, but compare with their own aspiration level.
If the payoffs do not achieve this aspiration level, they switch their states~\cite{pereira_pavlov}.

We have performed numerical simulations, where each simulation consists on distributing the players in a one-dimensional lattice with $L$ agents, out of which $\rho_0 L$ are randomly set as cooperators and the remaining ones as defectors.
Then agents play the IPD with $z$ neighbors and update their states according to the adopted strategy.
This process is repeated till the system reaches a steady state.
To calculate the asymptotic proportion of cooperators mean value and its standard deviation, we repeat the simulation for $M = 1000$ realizations, with different initial configurations of cooperators for each parameters set.
We present the results of the parameter space exhaustive exploration, $\rho_\infty(T = 1;~\rho_0;~z)$, and the associated standard deviation, $SD$, for players adopting the DES and the PES.
The exhaustive exploration is done using the following ranges and steps: $0 \leq \rho_0 \leq 1$ with $\Delta\rho_0 = 0.1$ steps and $2 \leq z \leq 30$ with integer values.

The state switching processes drive the system to a steady state that defines a phase~\cite{pereira_pavlov,pereira_2008_IJMPC,pereira_2008_BJP}.
In a naive analysis, when the system exhibits a cooperators majority ($\rho_\infty > 0.5$), we say that the system is in the cooperative phase.
If a defectors majority occurs ($\rho_\infty < 0.5$), we have the defective phase.
Also, a chaotic phase can occur, characterized by a high sensitivity to small change in the initial configuration, leading to large $\rho_\infty$ fluctuations.
As it is shown in the Ref.~\cite{pereira_pavlov}, in the {\it quasi}-regular phase, the proportion of cooperators oscillates around $\rho_\infty \sim 0.5$, with many players switching their states at each time step, even in the steady state.
In cooperative, defective and {\it quasi}-regular phases, the $\rho_\infty$ fluctuations (standard deviation) are very small, contrasting to the chaotic ones.
For a system, which players adopt the DES, the cooperative, defective or chaotic phases may occur; if players adopt the PES, the possible outcomes are: the cooperative or {\it quasi}-regular phases~\cite{pereira_pavlov}.

\begin{figure}[!htbp]
\centering
\resizebox{0.5\columnwidth}{!}{
	\includegraphics{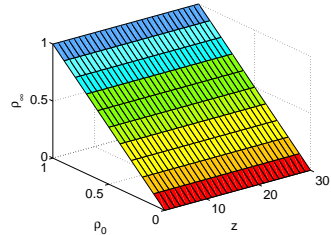}
}
\caption{
If $T = 1$ was a trivial case, the $\rho_\infty$ surface, as function of $\rho_0$ and $z$, is not altered from the initial configuration due time evolution.
}
\label{fig_T1}
\end{figure}

To compare our results, consider $T = 1$ as a trivial case.
As currently believed, in this case, all players do not switch their states during time evolution and the system keeps its initial configuration 
($\rho_\infty = \rho_0$) for every parameters set.
For $T = 1$, the $\rho_\infty$ surface, as function of $\rho_0$ and $z$, is a plan depicted in Fig.~\ref{fig_T1}.

\begin{figure}[!htbp]
\subfloat[]{
\label{fig_evo_T1_superficie_par}
\resizebox{0.45\columnwidth}{!}{
	\includegraphics{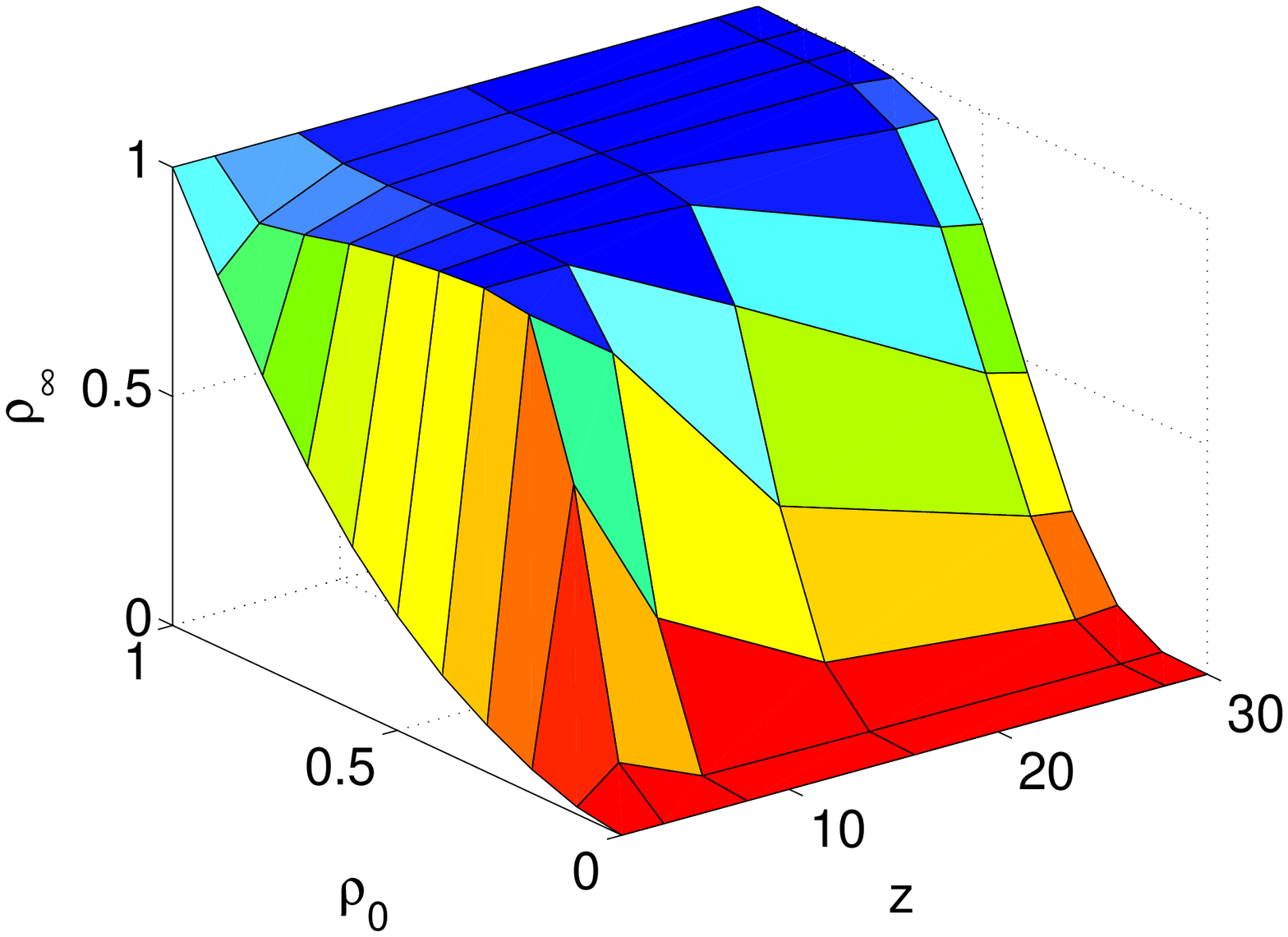}}
}
\subfloat[]{
\label{fig_evo_T1_superficie_impar}
\resizebox{0.45\columnwidth}{!}{
	\includegraphics{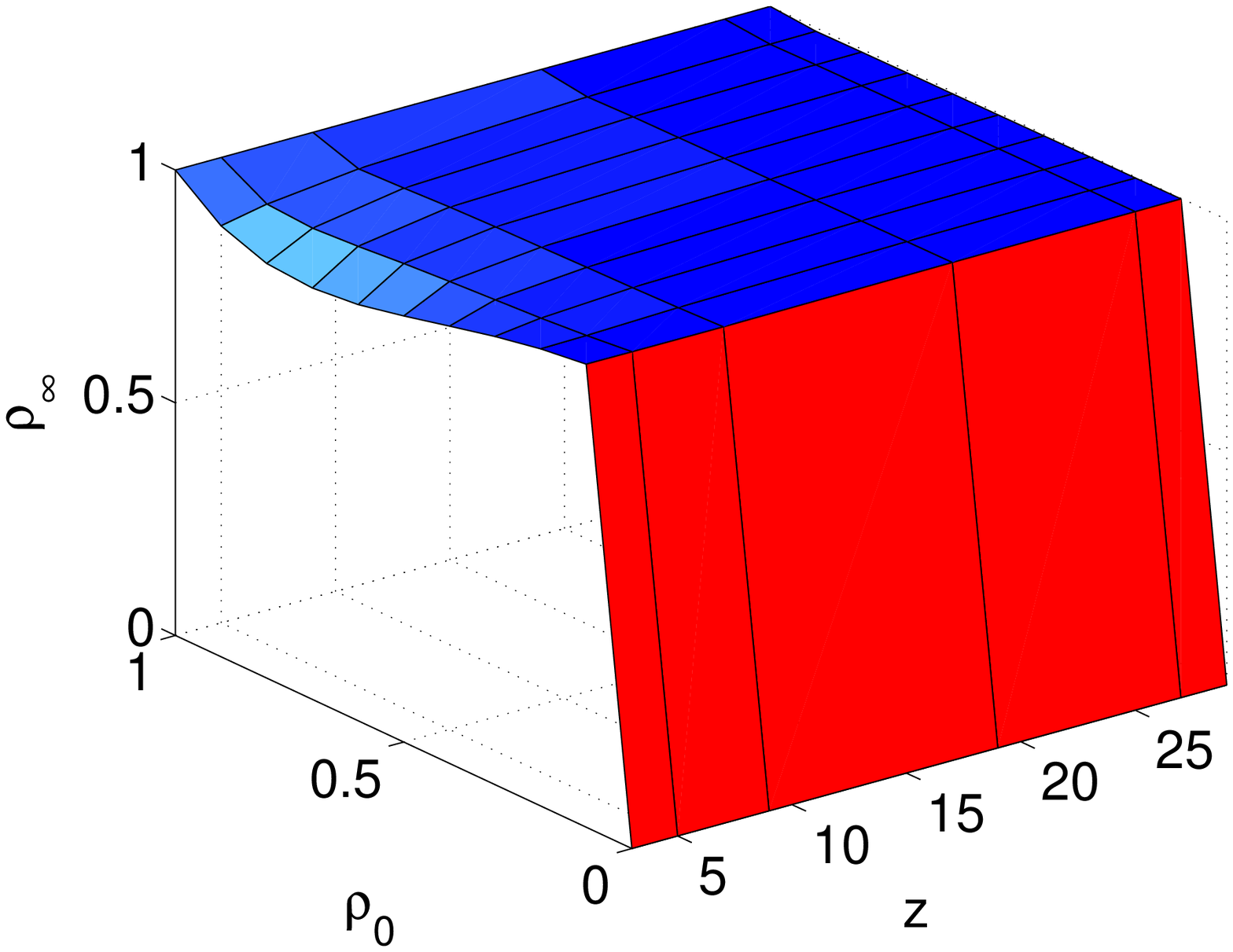}} 
} \\
\subfloat[]{
\label{fig_evo_T1_superficie_par_SD}
\resizebox{0.45\columnwidth}{!}{
	\includegraphics{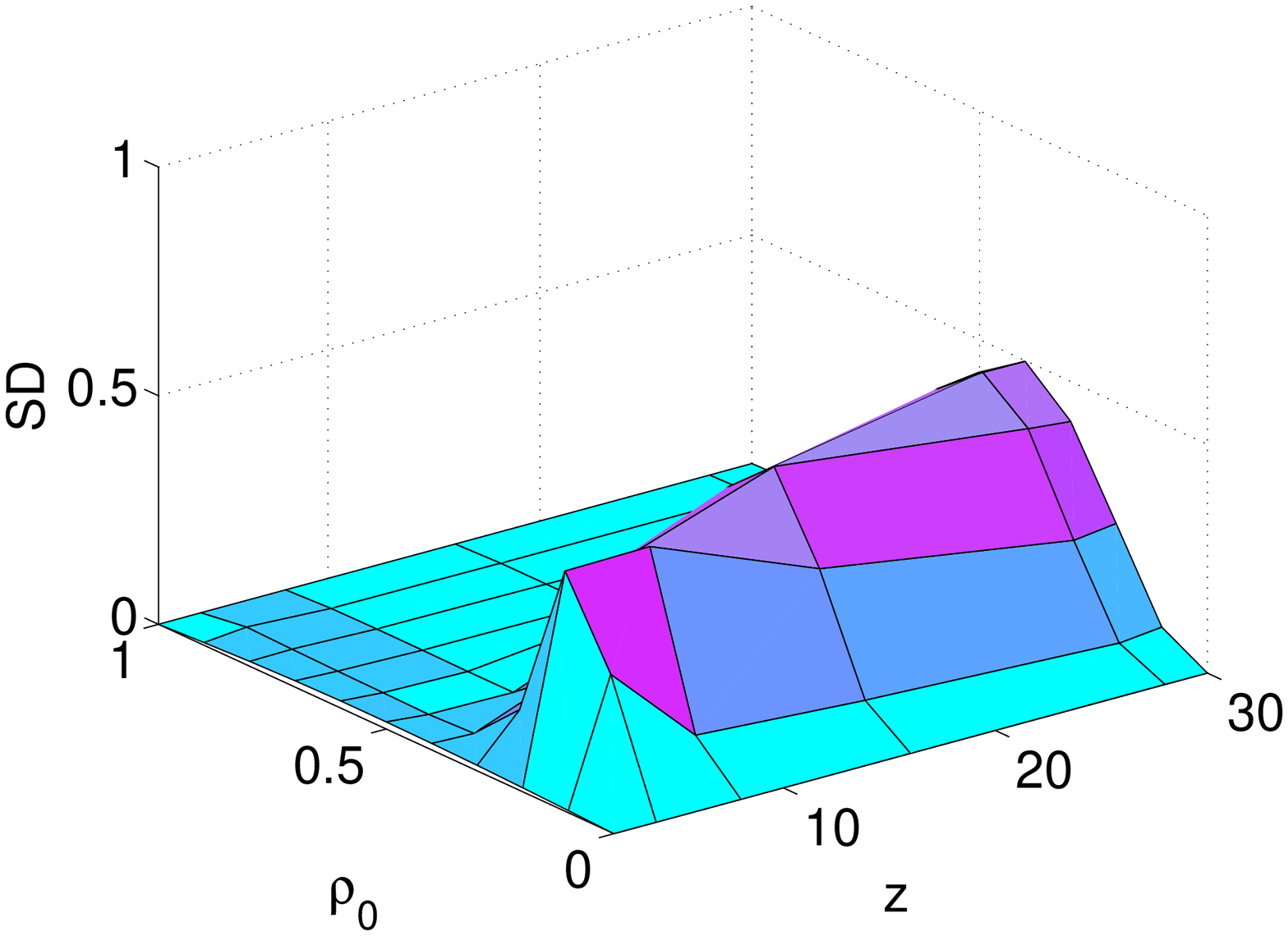}}
}
\subfloat[]{
\label{fig_evo_T1_superficie_impar_SD}
\resizebox{0.45\columnwidth}{!}{
	\includegraphics{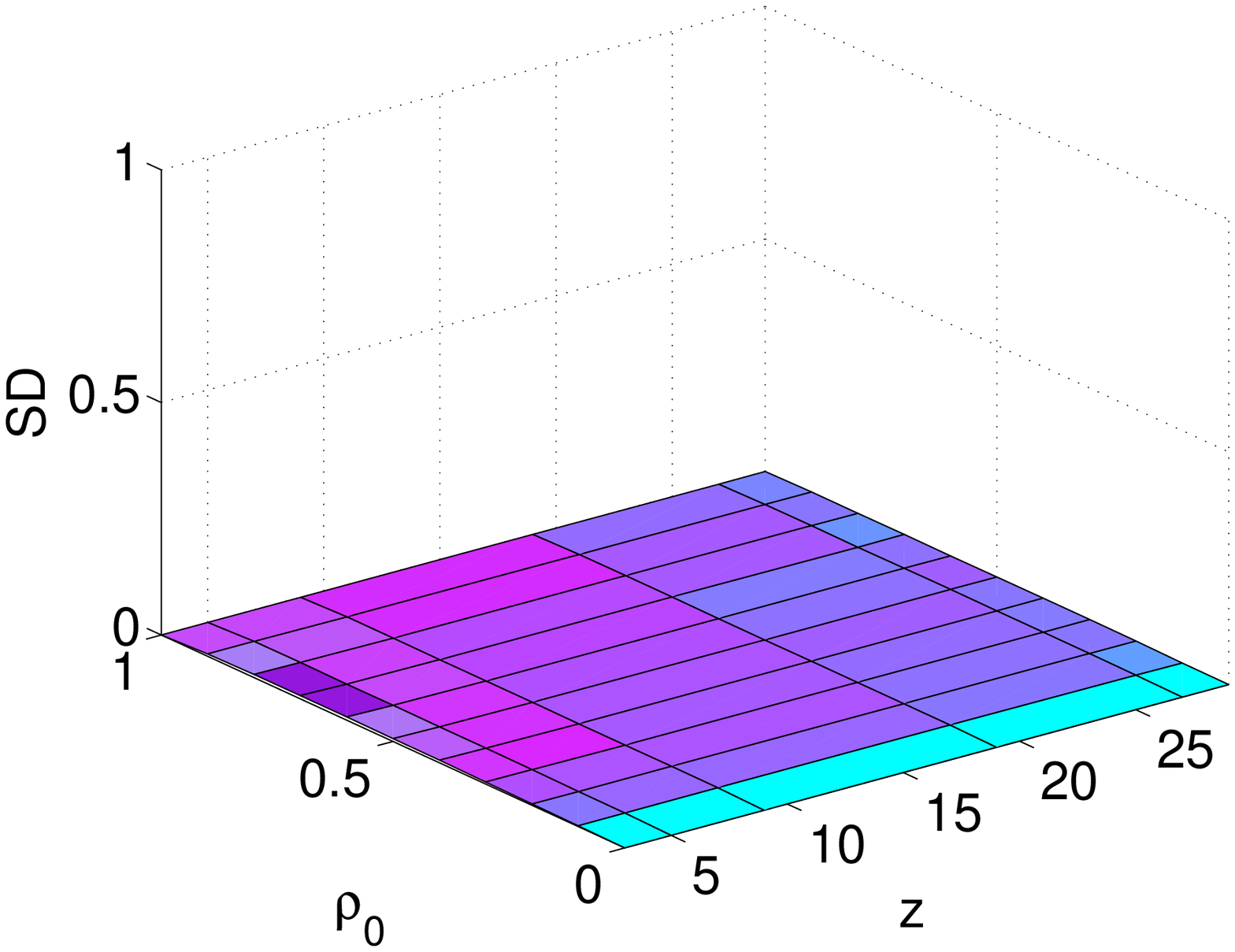}}
} \\
\subfloat[]{
\label{fig_evo_T1_p0_z_par}
\resizebox{0.45\columnwidth}{!}{
	\includegraphics{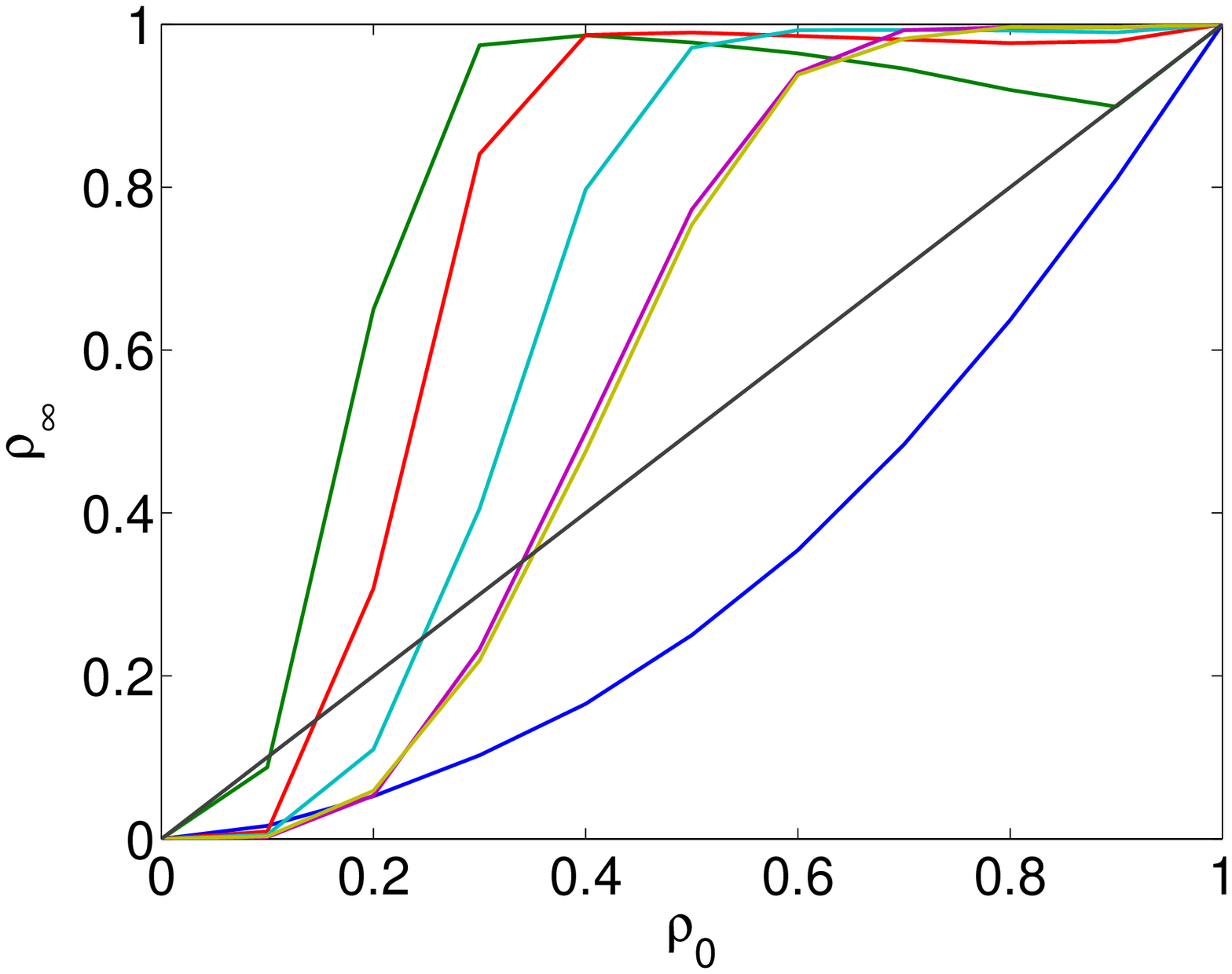}}
}
\subfloat[]{
\label{fig_evo_T1_p0_z_impar}
\resizebox{0.45\columnwidth}{!}{
	\includegraphics{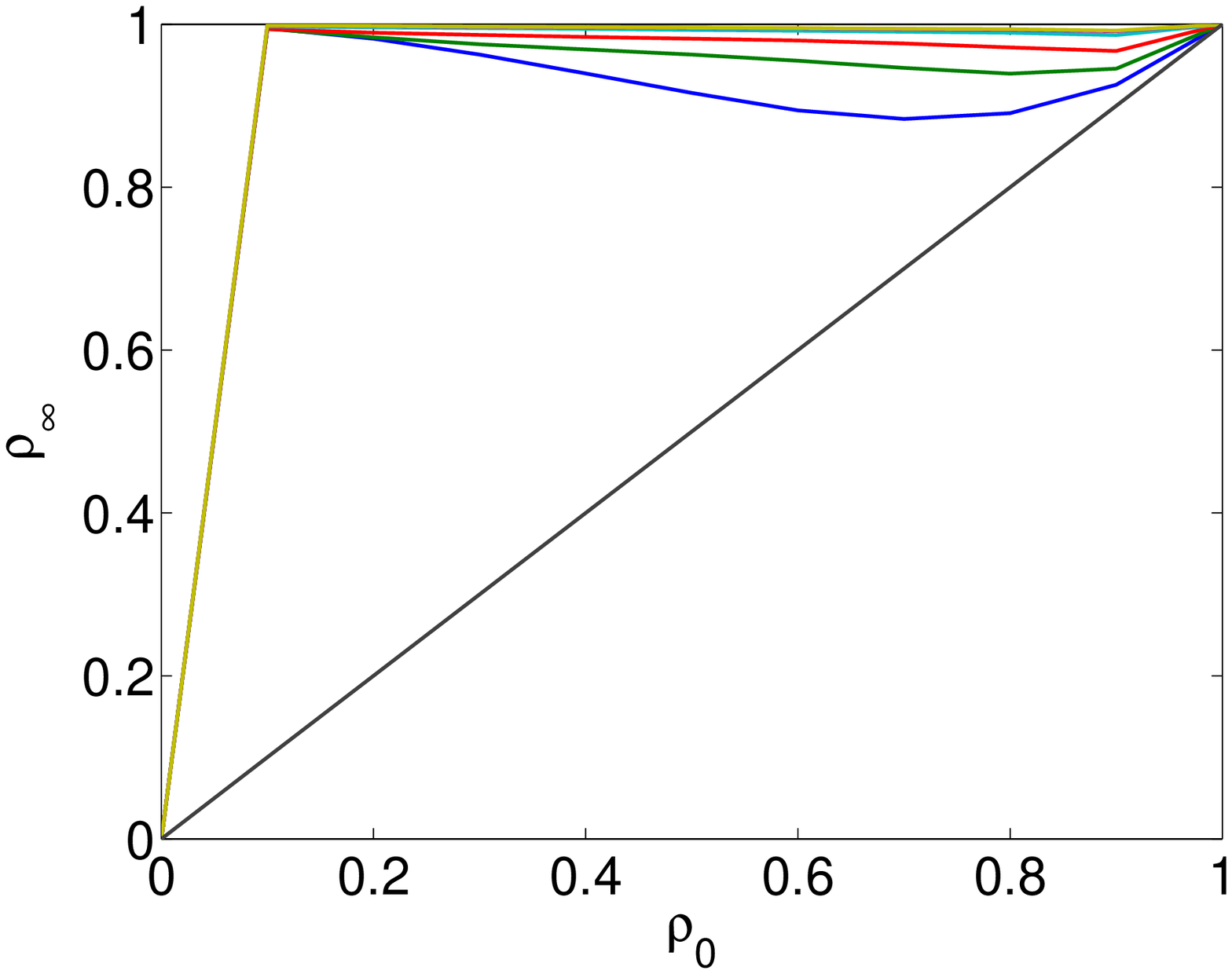}}
}
\caption{
Surfaces of $\rho_\infty$ and their fluctuations as a function of $\rho_0$ and $z$, for $T = 1$.
Players adopt the DES.
Specifically, (a) $\rho_\infty$ for even $z$ values, with $z = [2;~30]$ (without self-interaction);
(b) $\rho_\infty$ for odd $z$ values, with $z = [3;~29]$ (with self-interaction);
(c) fluctuation of $\rho_\infty$ displayed in (a);
(d) the same for (b).
Projections of $\rho_\infty$ in the plan $\rho_\infty \rho_0$:
(e) blue: $z = 2$, green: $z = 4$, red: $z = 8$, cyan: $z = 16$, magenta: $z = 28$, yellow: $z = 30$, black: result if players did not change their states.
(f) blue: $z = 3$, green: $z = 5$, red: $z = 9$, cyan: $z = 19$, magenta: $z = 27$, yellow: $z = 29$, black color: trivial case.
}
\label{fig_evo_T1}
\end{figure}

The $\rho_\infty(T = 1;~\rho_0;~z)$ surfaces, for DES, and its standard deviation are depicted in Fig.
\ref{fig_evo_T1}.
Figs. 
\ref{fig_evo_T1_superficie_par} (without self-interaction, even $z$ values) and 
\ref{fig_evo_T1_superficie_impar} (with self-interaction, odd $z$ values) show that players switch theirs states, because $\rho_\infty$ does not form the plan as the one displayed in Fig.~\ref{fig_T1}.
Players have not kept their initial states during system evolution.
Players switch their states, because they compare the total payoff, not the payoff per play of each pair of players.
Observing the systems, with and without self-interaction, one can notice that the cooperative phase is more prominent than the defective one in both cases.
However, self-interaction increases the cooperative phase in comparison to the absence of self-interaction.
Once the players self-interact, a cooperator has at least one positive payoff and a defector has a null payoff.
In this way, self-interaction is advantageous to cooperator and they can be replicated more easily (higher payoff) and cooperation emerges in the system.
Figs.
\ref{fig_evo_T1_superficie_par_SD} and \ref{fig_evo_T1_superficie_impar_SD} show the graphics of the fluctuations of $\rho_\infty$ for even and odd $z$ values, respectively.
In these graphics, if $SD \sim 0.5$, the system presents the chaotic phase in that region of the parameter space.
For $T = 1$, the chaotic phase is present only for even $z$ values, without self-interaction (see Fig.~\ref{fig_evo_T1_superficie_par_SD}).
The chaotic phase occurs between the cooperative and defective phases (see the cliff in Fig.~\ref{fig_evo_T1_superficie_par}) as $\rho_0$ decreases.
When players self-interact, cooperation increases and the chaotic phase does not appear at all(see Fig.~\ref{fig_evo_T1_superficie_impar_SD}).

The projection of $\rho_\infty(1;~\rho_0;~z)$ on the plan $\rho_\infty \rho_0$ represents $\rho_\infty$ as a function of $\rho_0$ for different $z$ values.
Fig.
\ref{fig_evo_T1_p0_z_par} displays these projections for even $z$ values (without self-interaction) and Fig.
\ref{fig_evo_T1_p0_z_impar}, the projections for odd ones (with self-interaction).
In Fig. 
\ref{fig_evo_T1_p0_z_impar}, the cooperation emerges for $\rho_0 \geq 0.1$, due the self-interaction presence, which increases the cooperators payoff, as explained above.
In Fig. 
\ref{fig_evo_T1_p0_z_par}, $\rho_\infty$ values can be greater or lower than the ``expected'' value: $\rho_\infty = \rho_0$.
For $\rho_0 < 0.4$, $\rho_\infty < \rho_0$ and $SD \sim 0.5$.
On one hand, if the cooperator proportion is small ($\rho_0 \sim 0$), these cooperators receive several null payoffs (due to the interactions with defectors) and their payoffs decrease, while the defector payoffs increase, consequently cooperators switch their states and cooperation does not emerge.
On the other hand, when $\rho_0 > 0.4$ more cooperators exist in the system in the beginning of the time evolution.
Cooperators playing against themselves receive a positive payoff.
The total payoff of cooperators become greater than the one of defectors that confronted others defectors, so these cooperators do not switch.
But, the defectors, despite of exploiting these cooperators, can confront others defectors, which lead to a decrease of their total payoff, leading them to switch their states, which drive the system to the defective phase.
Meanwhile, for $z = 2$, it occurs one exception, in this case the cooperation does not emerge, because cooperator $j$, that plays against cooperator $i$ and defector $k$, has a payoff $G_j = 1$.
If defector $k$ interacts with other cooperator beyond $j$ (remember that $z = 2$), he/she has a payoff $G_k = 2$, then the cooperator $j$ copies the player $k$ state.
In this way, defection becomes the main behavior of the players, raising the defective phase.
This allows us to conclude that higher connectivity favors cooperation in this case, because it increases the chance of a cooperator to interact with other cooperators.

\begin{figure}[!htbp]
\subfloat[]{
\label{fig_pav_T1_superficie_par}
\resizebox{0.45\columnwidth}{!}{
	\includegraphics{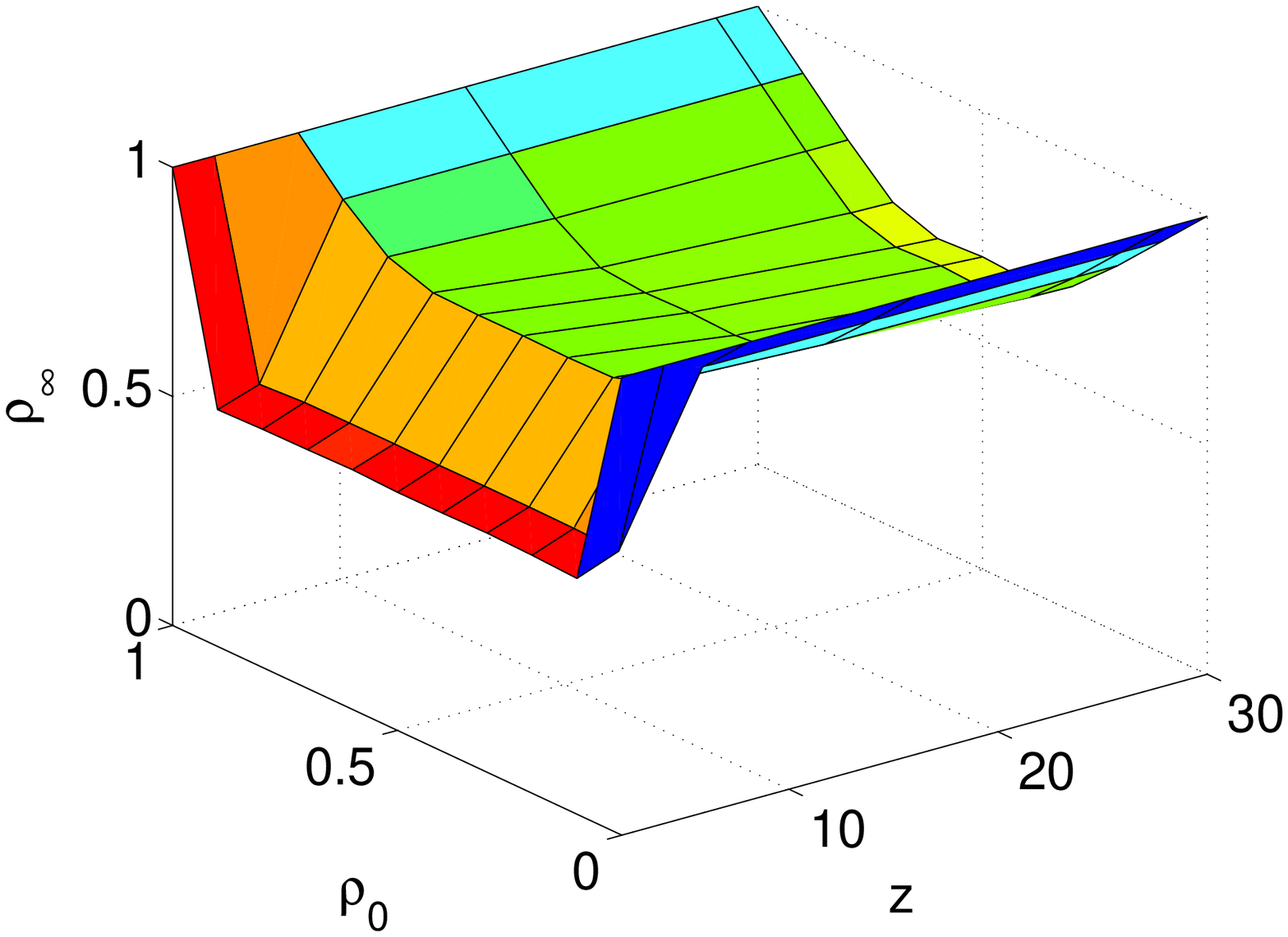}}
}
\subfloat[]{
\label{fig_pav_T1_superficie_impar}
\resizebox{0.45\columnwidth}{!}{
	\includegraphics{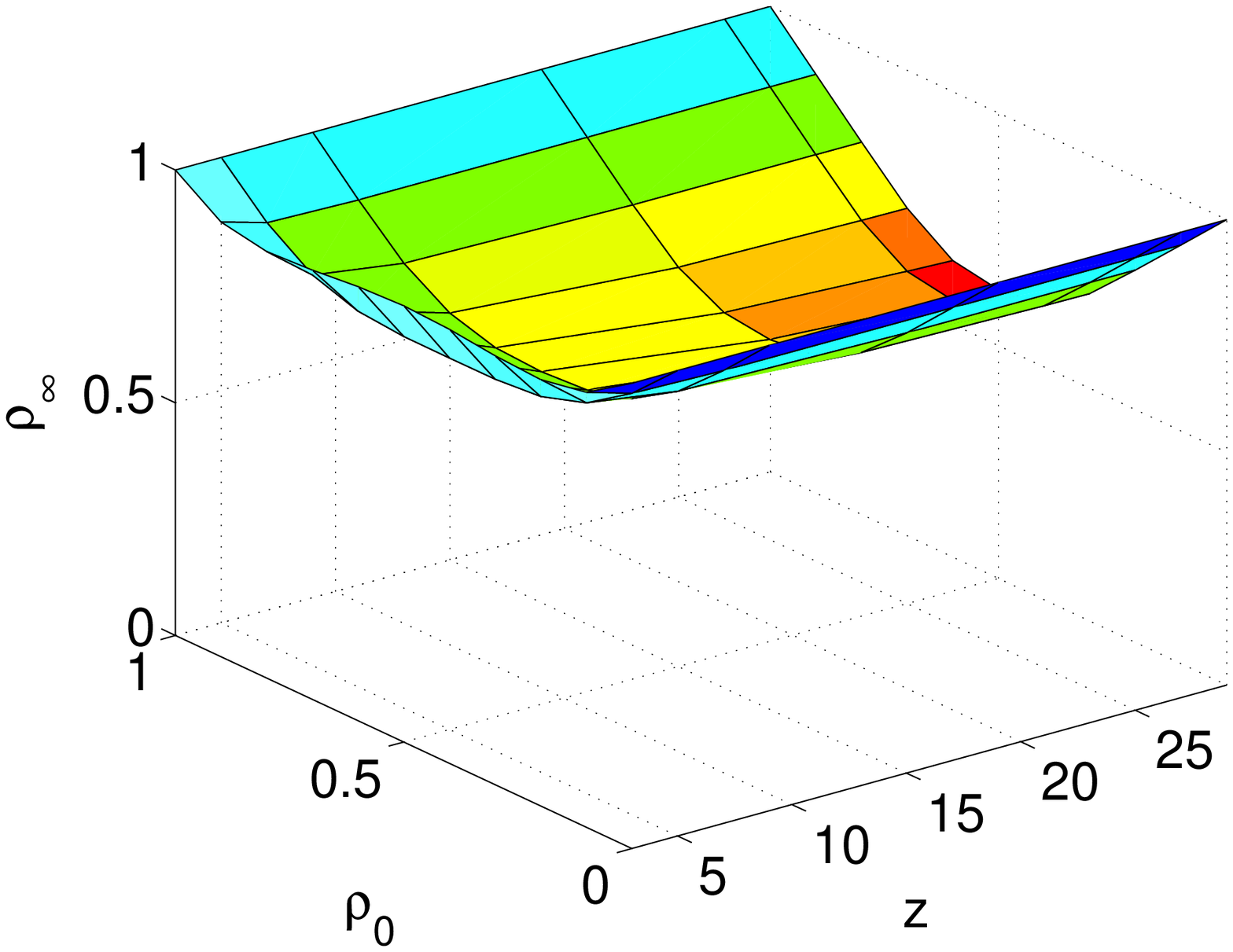}}
} \\
\subfloat[]{
\label{fig_pav_T1_superficie_par_SD}
\resizebox{0.45\columnwidth}{!}{
	\includegraphics{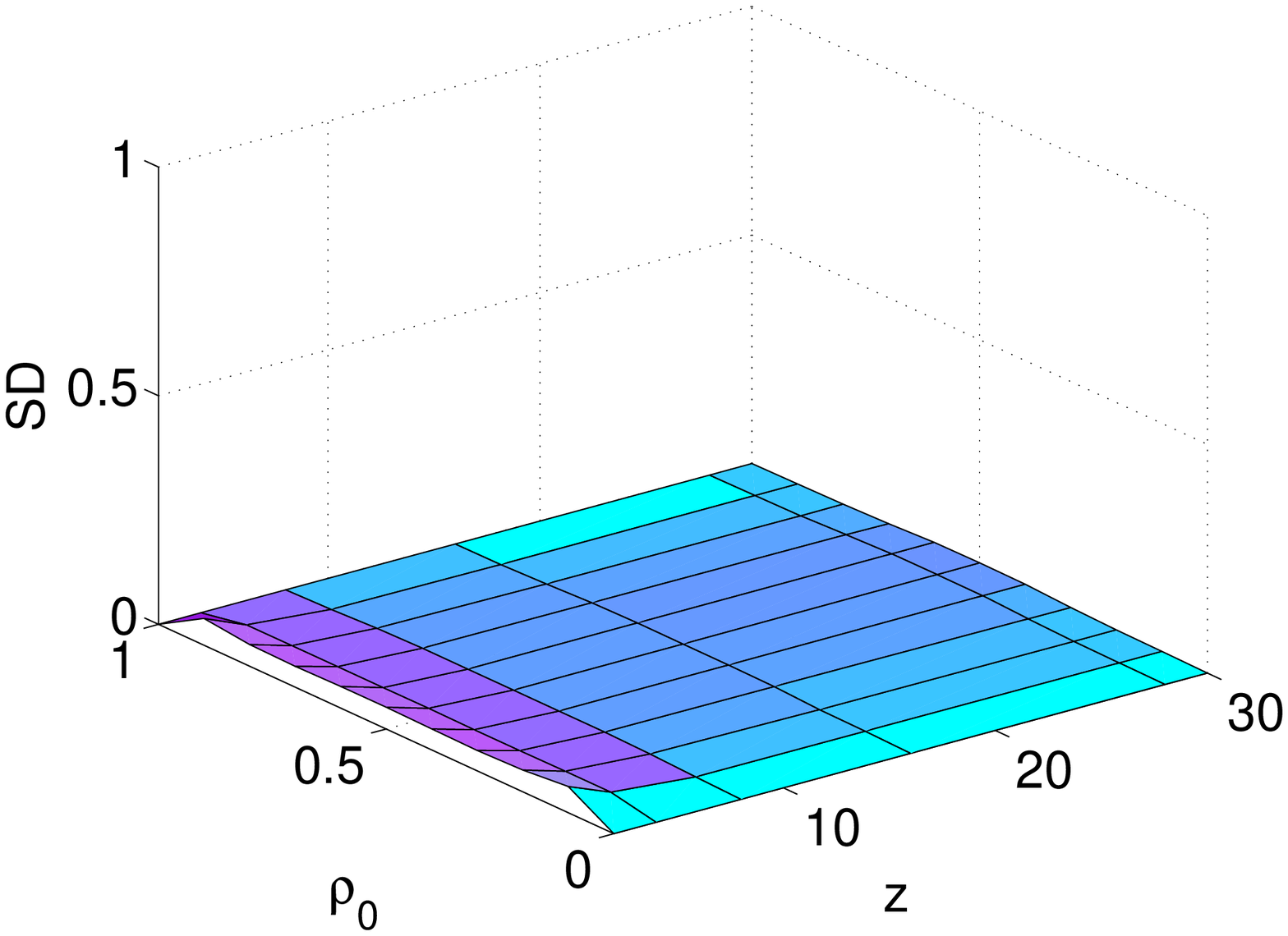}}
}
\subfloat[]{
\label{fig_pav_T1_superficie_impar_SD}
\resizebox{0.45\columnwidth}{!}{
\includegraphics{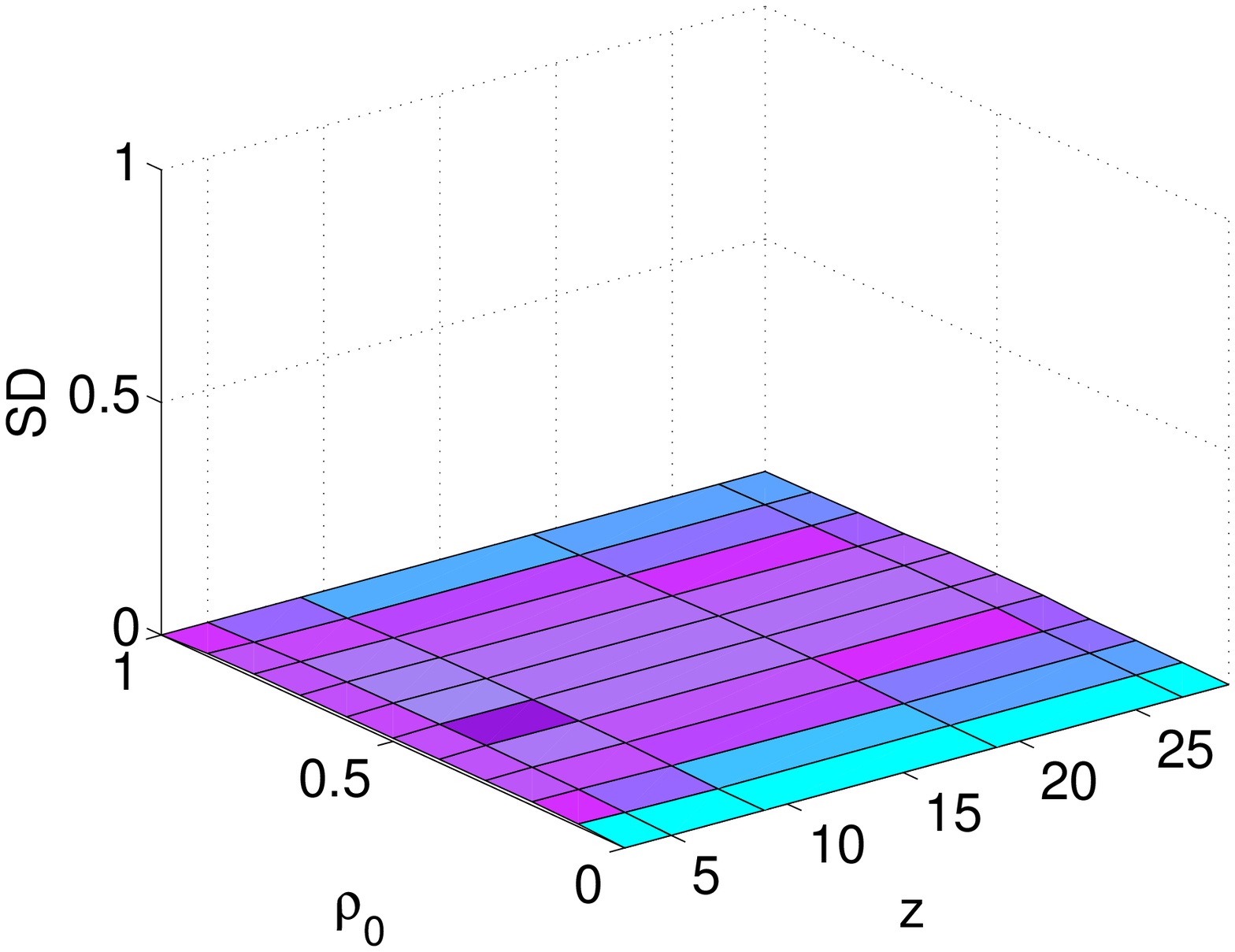}}
} \\
\subfloat[]{
\label{fig_pav_T1_p0_z_par}
\resizebox{0.45\columnwidth}{!}{
	\includegraphics{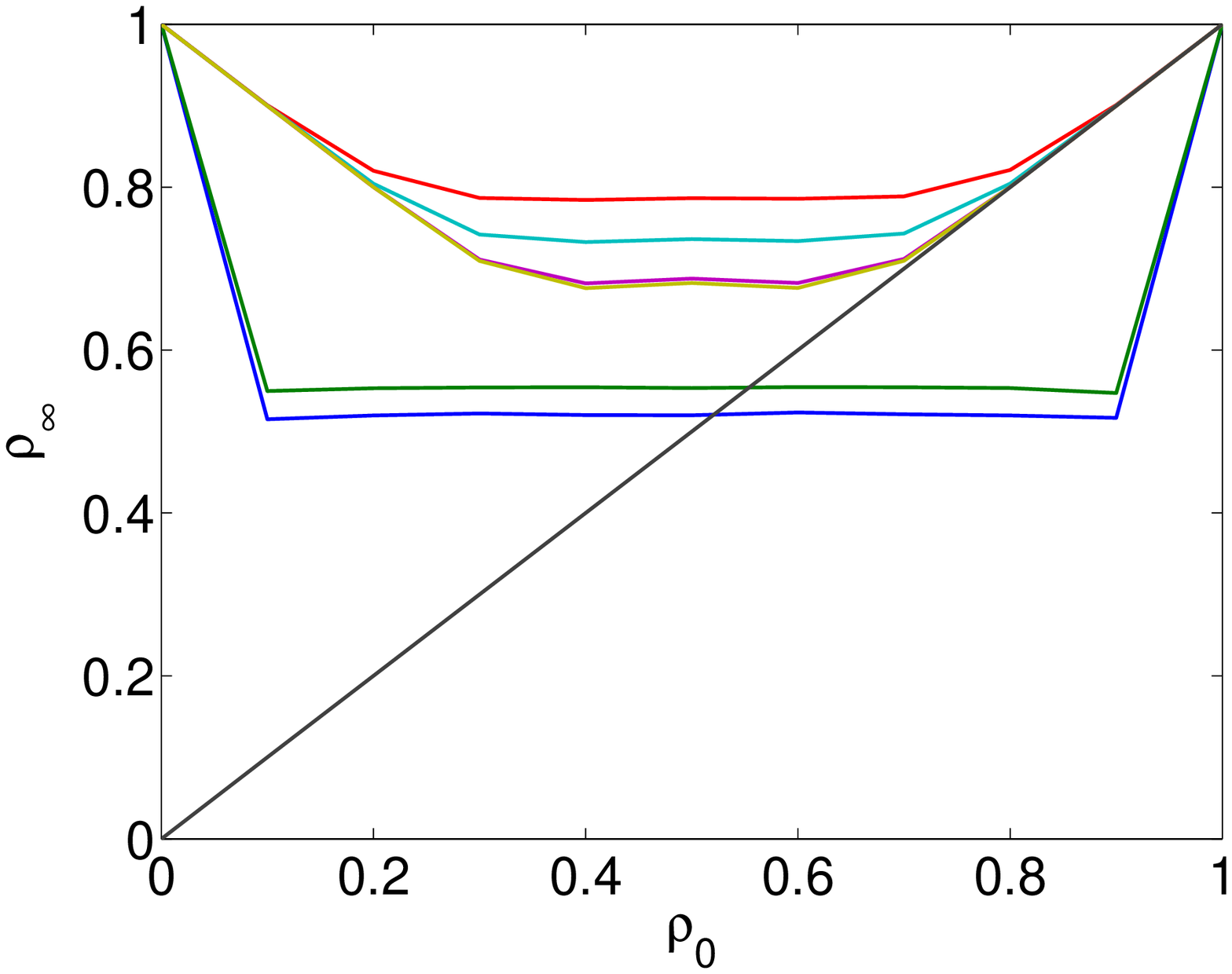}}
}
\subfloat[]{
\label{fig_pav_T1_p0_z_impar}
\resizebox{0.45\columnwidth}{!}{
	\includegraphics{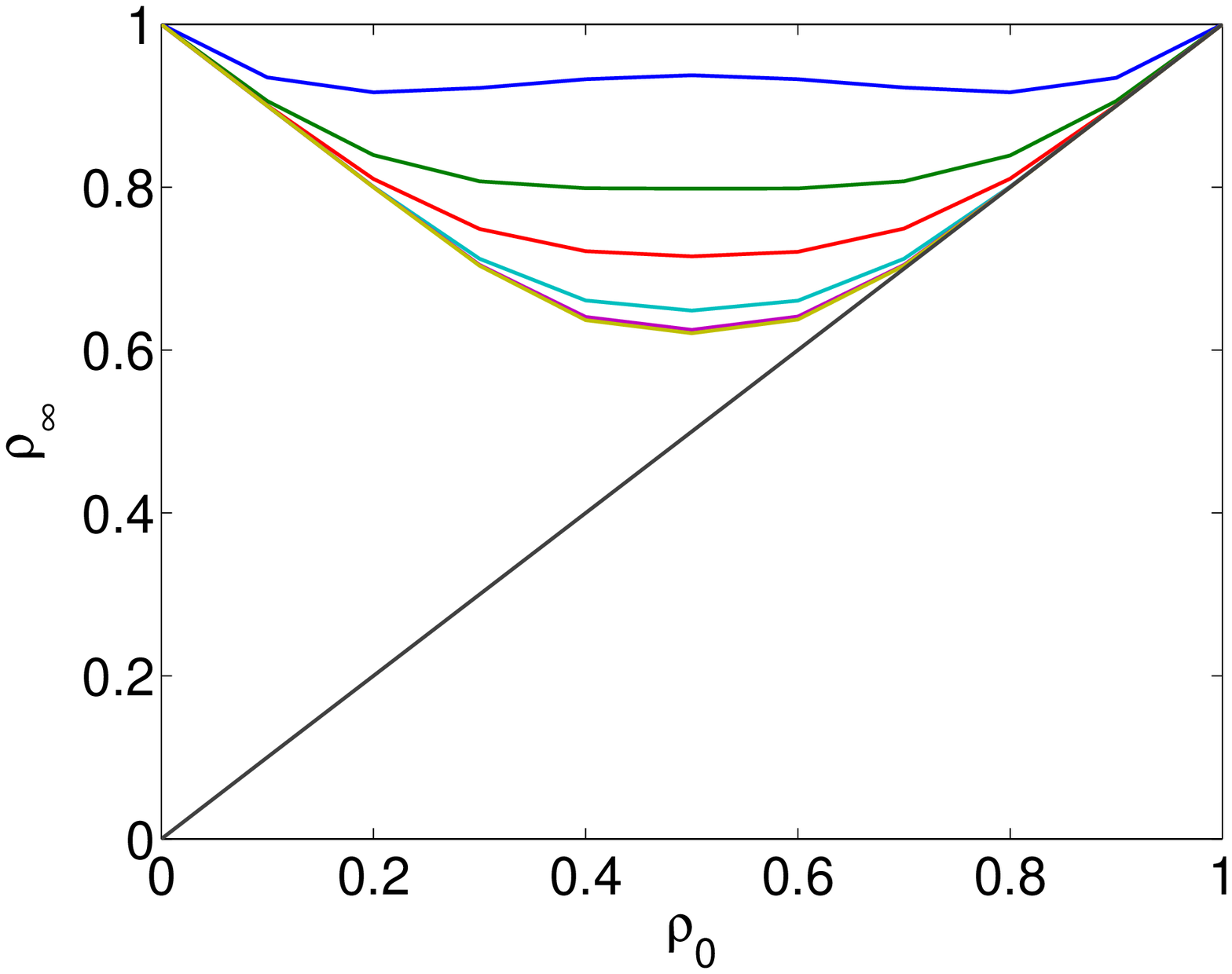}}
}
\caption{
Surfaces of $\rho_\infty$ and their fluctuations as a function of $\rho_0$ and $z$, for $T = 1$.
Players adopt the PES.
Specifically, (a) $\rho_\infty$ for even $z$ values, with $z = [2;~30]$ (without self-interaction);
(b) $\rho_\infty$ for odd $z$ values, with $z = [3;~29]$ (with self-interaction);
(c) fluctuation of $\rho_\infty$ displayed in (a);
(d) the same for (b).
Projections of $\rho_\infty$ in the plan $\rho_\infty \rho_0$:
(e) blue: $z = 2$, green: $z = 4$, red: $z = 8$, cyan: $z = 16$, magenta: $z = 28$, yellow: $z = 30$, black: result if players did not change their states.
(f) blue: $z = 3$, green: $z = 5$, red: $z = 9$, cyan: $z = 19$, magenta: $z = 27$, yellow: $z = 29$, black color: trivial case.
}
\label{fig_pav_T1}
\end{figure}

For the PES, the payoff of two defectors, when playing against themselves, is negative ($P = -T = -1$).
This is enough for both to switch their states, but they still can explore cooperators of their neighborhoods.
If even exploiting the neighbors, their payoffs do not become positive, they switch their states and cooperation emerges in the system.
The $\rho_\infty(T = 1;~\rho_0;~z)$ surfaces, for PES, and the associated standard deviation are depicted in Fig.
\ref{fig_pav_T1}.
Figs. 
\ref{fig_pav_T1_superficie_par} (without self-interaction, even $z$ values) and 
\ref{fig_pav_T1_superficie_impar} (with self-interaction, odd $z$ values) are very different from those observed for the DES (Fig.
\ref{fig_evo_T1_superficie_par} and \ref{fig_evo_T1_superficie_impar}).
Nevertheless, they also confirm that proportion of cooperators evolves as time goes by, for $T = 1$.
Here, the majority of the players cooperate in whole the parameter space with the exception of $z = 2$ and $z = 4$ (without self-interaction, see Fig.~\ref{fig_pav_T1_superficie_par}), where the {\it quasi}-regular phase occurs, with $\rho_\infty \sim 0.5$.
Notice that $\rho_\infty$ decreases as $z$ increases, with a stiffness more pronounced in the presence of self-interaction, because defectors always receive a null payoff due his/her self-interaction.
Also, a $\rho_\infty$ symmetry occurs, regarding the $\rho_0 = 1/2$, because for $\rho_0 = \beta$, while $rL$ players receive a positive payoff, $(1-r)L$ players receives a negative one, whether, $\rho_0 = 1 - \beta$, $(1-r)L$ players receive a positive payoff, $rL$ players receives a negative one, where $r$ is a arbitrary proportion of players which depends on the distribution of the players in each time step~\cite{pereira_pavlov}.
Figs.
\ref{fig_pav_T1_superficie_par_SD} and \ref{fig_pav_T1_superficie_impar_SD} show the $SD$ for even and odd $z$ values, respectively.
In these graphics, the small $SD$ values show the non-existence of the chaotic phase.
Only the {\it quasi}-regular ($z = \{2;~4\}$) phase has a slightly higher value.
Fig.
\ref{fig_pav_T1_p0_z_par} displays the projections of $\rho_\infty$ on the plan $\rho_\infty \rho_0$ for even $z$ values (without self-interaction), and Fig.
\ref{fig_pav_T1_p0_z_impar} for odd ones (with self-interaction).
These plots demonstrate that the cooperative phase is dominant.
The {\it quasi}-regular phase occurs only for $z = 2$ and $z = 4$.
The proportion of cooperators is below than the ``expected'' only in the {\it quasi}-regular phase.

The results presented here show that, for $T = 1$, the system is not static and trivial, as previously supposed, when players play the IPD with more than one neighbor ($z > 1$).
On one hand, if players adopt the DES, cooperative, defective and chaotic phases may be present.
The chaotic phase appears only for even $z$ values (without self-interaction).
The more astonishing result is the cooperative phase ($\rho_\infty \sim 1$), that is present for $\rho_0 > 0.5$, without self-interaction and for $\rho_0 > 0$ with it.
On the other hand, adopting the PES, as $z$ increases, $\rho_\infty(1;~\rho;~z \gg 1)$ decreases with the exception $z = \{2;~4\}$, with a decrease more pronounced when the self-interaction is present.
For $z = \{2;~4\}$, system presents the {\it quasi}-regular phase, which we have firstly pointed out in Ref.~\cite{pereira_pavlov}.
The symmetry of $\rho_\infty$ can be explained by equivalence arguments.
Cooperation emerges even when cooperators and defectors have the same payoff in the IPD.
The increase in the connectivity favors cooperation for the DES, but it decreases the cooperation for the PES.
The initial proportion of cooperators is not a relevant parameter in this problem, it has less influence in the results than the others parameters.

Finally, the authors acknowledge that they have greatly profited from the discussions with H. Fort and R. da Silva.
M. A. P. thanks CAPES and CNPq for fellowships.
A. S. M. acknowledges the support of CNPq (303990/2007-4, 476862/2007-8) and CNPq/PROSUL Project (490440/2007-0).

\bibliographystyle{unsrt}
\bibliography{T1_arxiv.bbl}

\begin{thebibliography}{10}

\bibitem{axelrod_1984}
R.~Axelrod.
\newblock {\em The evolution of cooperation}.
\newblock Basic Books, New York, 1984.

\bibitem{axelrod_1980}
R.~Axelrod.
\newblock Effective choice in the prisoner's dilemma.
\newblock {\em J. Confl. Resolut.}, 24:3--25, 1980.

\bibitem{axelrod_1980b}
R.~Axelrod.
\newblock More effective choice in the prisoner's dilemma.
\newblock {\em J. Confl. Resolut.}, 24:379--403, 1980.

\bibitem{axelrod_1981}
R.~Axelrod and W.~D. Hamilton.
\newblock The evolution of cooperation.
\newblock {\em Science}, 211:1390--1396, 1981.

\bibitem{nowak_1992}
M.~A. Nowak and R.M. May.
\newblock Evolutionary games and spatial chaos.
\newblock {\em Nature}, 359:826--829, 1992.

\bibitem{soares_2006}
R.~O.~S. Soares and A.~S. Martinez.
\newblock The geometrical patterns of cooperation evolution in the spatial
  prisoner's dilemma: An intra-group model.
\newblock {\em Physica A}, 369:823--829, 2006.

\bibitem{fort_2005}
H.~Fort and S.~Viola.
\newblock Spatial patterns and scale freedom in prisoner's dilemma cellular
  automata with pavlovian strategies.
\newblock {\em Journal Of Statistical Mechanics-Theory And Experiment},
  1:P01010, 2005.

\bibitem{kraines_1989}
D.~Kraines and V.~Kraines.
\newblock Pavlov and the prisoner's dilemma.
\newblock {\em Theory and Decision}, 26:47--79, 1989.

\bibitem{kraines_1993}
D.~Kraines and V.~Kraines.
\newblock Learning to cooperate with pavlov - an adaptive strategy for the
  iterated prisoners-dilemma with noise.
\newblock {\em Theory Decis.}, 35(2):107--150, 1993.

\bibitem{kraines_1995}
D.~Kraines and V.~Kraines.
\newblock Evolution of learning among pavlov strategies in a competitive
  environment with noise.
\newblock {\em J. Confl. Resolut.}, 39(3):439--466, 1993.

\bibitem{posch_1999}
M.~Posch.
\newblock Win-stay, lose-shift strategies for repeated games - memory lenght,
  aspiration levels and noise.
\newblock {\em J. theor. Biol.}, 198:183--195, 1999.

\bibitem{lorberbaum_2002}
J.~P. Lorberbaum, D.~E. Bohning, A.~Shastri, and L.~E. Sine.
\newblock Are there really no evolutionarily stable strategies in the iterated
  prisoner's dilemma?
\newblock {\em J. Theor. Biol.}, 214(2):155--169, 2002.

\bibitem{nowak_1993b}
M.~A. Nowak and K.~Sigmund.
\newblock A strategy of win-stay, lose-shift that outperforms tit-for-tat in
  the prisoner's dilemma game.
\newblock {\em Nature}, 364:56--58, 1993.

\bibitem{beyer_2002}
H.~G. Beyer and H.~P. Schwefel.
\newblock Evolution strategies.
\newblock {\em Natural Computing}, 1:3--52, 2002.

\bibitem{pereira_pavlov}
M.~A. Pereira and A.~S. Martinez.
\newblock Pavlovian prisoner's dilemma in one-dimensional cellular automata:
  analytical results, the {\it quasi}-regular phase, spatio-temporal patterns
  and parameter space exploration.
\newblock Arxiv: arXiv:0904.0384v1 [physics.soc-ph].

\bibitem{duran_2005}
O.~Dur\'{a}n and R.~Mulet.
\newblock Evolutionary prisoner's dilemma in random graphs.
\newblock {\em Physica D}, 208:257--265, 2005.

\bibitem{pereira_2008_IJMPC}
M.~A. Pereira, A.~S. Martinez, and A.~L. Esp\'{i}ndola.
\newblock Prisoner's dilemma in one-dimensional cellular automata:
  visualization of evolutionary patterns.
\newblock {\em Int. J. of Modern Phys. C}, 1, 2008.

\bibitem{pereira_2008_BJP}
M.~A. Pereira, A.~S. Martinez, and A.~L. Esp\'{i}ndola.
\newblock Exhaustive exploration of prisoner's dilemma parameter space in
  one-dimensional cellular automata.
\newblock {\em Braz. J. of Phys.}, 38(1):65--69, March.

\end{thebibliography}

\end{document}